\documentstyle[12pt]{article}
\def\ii{\'{\i}}
\def\bi{\bigskip}
\def\noi{\noindent}
\def\be{\begin{equation}}
\def\en{\end{equation}}
\def\bq{\begin{eqnarray}}
\def\eq{\end{eqnarray}}
\begin{document}

\begin{center}
{\Large \bf ON THE MAGNETO-ROTON GAP AND THE LAUGHLIN STATE STABILITY}
\\[1.5cm]
\end{center}
\begin{center}
{\large\bf Alejandro Cabo Montes de Oca\footnote{Work supported by CONACyT under contract 3979pe-9608}}\\[.3cm]
{\it Instituto de F\'\i sica, Universidad de Guanajuato}\\
{\it Lomas del Bosque \# 103, Lomas del Campestre}\\
{\it 37150 Le\'on, Guanajuato; M\'exico}\\
{\it and}\\
{\it Instituto de Cibern\'etica Matem\'atica y F\ii sica}\\
{\it Calle E, No. 309. Esq. a15, Vedado, La Habana, Cuba}
\end{center}
\noindent

\vspace{2.5cm}

\begin{center}
{\bf Abstract.} 
\end{center}

It is argued that the Girvin, MacDonald and Platzman (GMP) evaluation of 
the magneto-roton spectrum, in spite of probably being a sensible estimate
of the excitation spectrum around the real FQHE ground state, is not implying
the variational stability of the Laughlin state. The suplementary corrections 
needed to produce a variational calculation around the $\nu= 1/3$ Laughlin 
state are evaluated approximately. The results indicate within the considered 
aproximation, the existence of lower lying states for a range of wavevector 
values.

\setlength{\baselineskip}{1\baselineskip}

\newpage

\section{Introduction.} 

\bi
\bi 

  In a previous letter [1] the violation of a stability condition by the 
Single Mode Approximation for the evaluation of the excitation gap in 
FQHE ground state was argued. In other related works, the calculations of the 
collective mode dispersion, first for composite fermions in [2] and after from a
Behe-Salpeter approach based in a phenomenological ansatz for the
electron progrator [3], gave similar indications for the existence of lower 
energy states with broken translational invariance.

\bi

It should be expressed that attempts to give foundation for a broken 
translational symmetry in the FQHE ground state have been existing since the 
times of the discovery of the effect. Some of them can be traced out in Refs.
[5]-[12]. Concretely, we have been working in a particular direction of 
thinking which based in the obtention of exact Hartree-Fock solution at 
fractional filling factors [13]-[17]. The existence and interesting properties
of these solutions have been the main motivation for the expectation of
a broken symmetry ground state.   

\bi

In the present letter, the GMP evaluation of the magneto-roton spectrum is 
conceptually examined [18]. The aim is to precise its implications in 
connection with the variational stability of the Laughlin state. It is 
concluded that the GMP evaluation, while presumably being a valuable estimate
of the real excitation spectrum around the true ground state, is not implying 
that magneto-roton states have greater energies than the Laughlin ones. 
Complementary terms should be also calculated in order to verify the Laughlin 
state variational stability. Their evaluation under the use of the approximate
reduced density matrices given in [19], gives results signaling the existence 
of magneto-roton wave functions having lower energies than the Laughlin state
at $\nu= 1/3$.

\bi

To start, let us consider the difference between the magneto-roton  and 
Laughlin state energies

\be
\epsilon (k) = <\phi_k |H| \phi_k>-<\phi_L |H| \phi_L>
\en

\bi

\noi
where the magneto-roton state is given as usual by

\be
|\phi_k> =\frac{1}{\sqrt{N}} \rho_k |\phi_L>
\en

\bi

\noi and its norm is given by the projected static structure factor

\be
s(k) =<\phi_k |\phi_k> =\frac{1}{N} <\phi_L | \rho_k^+ \rho_k |\phi_L >
\en

\noi where $N$ is the number of particles.

\bi

Here, all the conventions for the definitions of the density operators, 
Hamiltonian, etc. are the ones given in Ref. [20]. More specifically $z=x+iy$ 
is the complex representation of a 2D-position vector and $k=k_x +i k_y$ (or
$q= q_x +i q_y$) the same representation for wavevectors. The magnetic field 
is taken along the negative $z$ axis and the magnetic length is set equal to 
one. The symmetric gauge is also assumed. The projected electron Hamiltonian
is given by

\be
H=\frac{1}{2} \int \frac{d^2 q}{(2\pi)^2} \upsilon (\vec q) (\rho^+_q \rho_q 
-\rho e^{-q q^*/2})
\en 

\bi

\noi with $\rho$ being the density of the state and the Coulomb potential $\upsilon
(\vec q) =2\pi/|\vec q|$. Finally, the projected density operator is
given by

\be
\rho_k = \sum^{N}_{j=1} exp \left[ -ik \frac{\partial}{\partial z_j} \right]
exp \left[ -\frac{ik^*}{2} z_j \right].
\en 

\bi

After some algebraical transformations (1) can be written in form

\bq
\epsilon(k) &=& \frac{<\phi_L | \rho^+_k [H,\rho_k ]|\phi_L >}{<\phi_L | \rho^+_k 
\rho_k | \phi_L>} \nonumber \\
&+& \frac{<\phi_L | \rho^+_k \rho_k H | \phi_L>-<\phi_L | \rho^+_k \rho_k |
\phi_L><\phi_L |H| \phi_L>}{<\phi_L |\rho^+_k \rho_k | \phi_L >} 
\eq
  
Then, the reality of $\epsilon (k)=\epsilon^* (k)$ allows to write for
$\epsilon (k) =\frac{1}{2} (\epsilon (k) + \epsilon^* (k))$

\be
\epsilon (k) = \Delta (k) +\delta (k),
\en

\noi where

\bi

\be
\Delta (k) = \frac{<\phi_L |[\rho^+_k, [H, \rho_k ]|\phi_L>}{2<\phi_L | 
\rho^+_k \rho_k | \phi_k >},
\en

\be
\delta (k)=\frac{\frac{1}{2} <\phi_L|H\rho^+_k \rho_k +\rho^+_k \rho_k H|\phi_L>-
<\phi_L|\rho^+_k \rho_k|\phi_L><\phi_L|H|\phi_L>}{<\phi_L | \rho^+_k \rho_k | \phi_L>} .
\en 

\bi

From relation (9) it follows that the formula used in Ref. [20] for the 
evaluation of $\Delta (k)$, the SMA mode energy, describe the excitation over
the Laughlin state $|\phi_L>$ whenever $|\phi_L>$ is an exact eigenstate of 
$H$. In such a case the expression (9) for $\delta (k)$, vanish identically. 
Therefore, in order to conclude that the evaluation of $\Delta (k)$ in Ref. 
[20] implies the variational stability of the Laughlin state it should be 
verified that $\delta (k)$ can be disregarded. This simple conclusion is the main 
point of this letter.

\bi

Here a first consideration of the above  question is presented. For this purpose
an evaluation of (9) was done by making use of the reduced density 
matrices given in Ref. [19]

\bq
\rho_M (z^\prime , z)=\Big(\frac{\nu}{2\pi}\Big)^M \prod^{M}_{k=1}\Big\{ exp
\Big(-\frac{z_k z^*_k}{4}- \frac{z^\prime_k z^{*\prime}_k}{4} + \frac{z_k
z^{\prime*}}{2} \Big) \Big\} \nonumber \\
\prod^{M}_{\scriptstyle i,j=1\atop
\scriptstyle i<j}  g((z^\prime_i-z^\prime_j)^* (z_i-z_j)) \,\,\, M=1,2,3,4 ;
\eq 

\noi where $z=(z_1,\ldots z_M), z^\prime = (z^\prime_1, \ldots z^\prime_M)$ and
$g(r^2)$ is the pair correlation function of the Laughlin state. It 
should be stressed that only for $M=1,2$ the density matrices in (10) are
almost exactly known. For $M=3,4$ the expresions (10) are approximate ones
obtained under the assumption that three and four point density matrices are
expressed as products of two-point correlation functions $g$. The use
of the density matrices (10) allows to calculate the following expression
for the correction $\delta (k)$

\bq
\delta (k)&=&\frac{1}{2s(k)}(\frac{\nu}{2\pi}) \int \frac{d^2 q}{(2\pi )^2} v
(\vec q)(\exp (k^* q)+\exp(kq^*))\cdot\int d^2x\, \exp (i(\vec k+\vec q)\vec x) g 
(\vec x^2_1) \nonumber \\
& &+\frac{1}{s(k)}\Big(\frac{\nu}{2\pi}\Big)^2\int\frac{d^2q}{(2\pi)^2} v (\vec q)
\Big( \exp (k^* q/2) + \exp (kq^*/2) \Big) \cdot \\
& &\int d^2 x_1 d^2 x_2 \exp \left( i\vec k\cdot\vec x_1+i\vec q\cdot \vec x_2 
\right) g (\vec x^2_1)g (\vec x^2_2) g((\vec x_1 -\vec x_2)^2) \nonumber \\  
& &+ \frac{1}{2s(k)}\cdot\Big(\frac{\nu}{2\pi}\Big)^3 \int \frac{d^2q}{(2\pi )^2} 
v(\vec q) \int d^2 x_1 d^2 x_2 d^2 x_3 \exp (i\vec k \cdot \vec x_1 +i \vec q 
\cdot \vec x_2) \nonumber \\
& &g(\vec x^2_1)g(\vec x^2_2) [ g(\vec x^2_3)g((\vec x_3 -\vec x_1)^2)g((\vec x_3 
-\vec x_2)^2)g((\vec x_3 -\vec x_2 -\vec x_1)^2)-1 ] \nonumber
\eq

\noi where $g(\vec x^2)$ is given by the acurrate analytical expression 
derived by Girvin

\be
g(\vec x^2)=1-\exp (-\vec x^2/2)+ \sum^\infty_{n=0} \frac{2}{(2n +1)} 
\Big(\frac{\vec x^2}{4} \Big)^{2n+1} C_{2n+1} \exp \Big( -\frac{\vec x^2}{4} \Big)
\en

\noi in which the coefficient $C_{2n+1}$ are reported in [10] for the 13 first 
integers and the values $\nu = 1/3$ and $\nu = 1/5$ for the filling factor.

\bi

The further evaluation of $\delta (k)$ through (11) was considered for the
$\nu =1/3$ state. In performing it, the Fourier transform of the Coulomb 
potential was regularized at long distances (small $\vec k$) in order that its
zero momentum component vanish, that is $v(\vec k=0)=0$. This procedure allows
to regularize and cancel singular terms associated to the non-decaying values 
of the pair correlation functions at infinity. The resulting finite terms 
are continuos upon the removal of the regularization. The numerical evaluation
was performed approximatelly using the Monte Carlo algarithm implemented in 
Mathemathica 3.0. The number of sample points was incremented up to a 
stabilization of the calculated values was noticed. The results for the 
$\epsilon (k) =\Delta (k)+\delta (k)$ are shown in Fig. 1 by the continuos 
curve. The  points correspond to the magneto-roton spectrun $\Delta (k)$.

\bi

The results in Fig.1 indicate, within the considered approximation, that the
magneto-roton states could have lower energies than the Laughlin states. This 
occurs for wavevector values in excess of $k r_0 \sim 1.5$.  At lower
values of $k$ the energy difference $\epsilon (k)$ tends to grow. This 
bahavior is similar to the one obtained in the previous work [2] for the
composite fermion excitation spectrum. In that case the growth reflected 
the tendency of the spectrum to reproduce the cyclotron resonances excitation
at low wavevectors. A similar picture was also obtained in the work [3].

\bi

We want to stress that the instability of Laughlin state suggested by the
present calculation does not invalidate the $\Delta (k)$ spectrum as an 
accurrate approximation for the exact collective mode. This statement is
supported by the fact that the Laughlin state is undoubtely a good
approximation for the exact ground state, then, the evaluation of the formula
(8) for $\Delta (k)$ corresponding to the exact increase in energy over  the
true ground state could effectively furnish good results  for the correct
gap. Therefore, the present argue in this letter should not be interpreted
as claiming the invalidation of the SMA approximation. The central point 
here supported is that the real ground state could be a weakly inhomogeneous
state not very much differing from the Laughlin wavefunction.

\bi

In summary, it is underlined that the magneto-roton spectrum evaluation is not
implying the variational the stability of the Laughlin state at $\nu =1/3$. 
The additional terms needed in checking such stability are approximately 
calculated. The results indicate the existence of magneto-roton states lowering
the energy of the $\nu =1/3$ Laughlin wavefunction. The work in performing 
precise evaluations needed to confirm the existence of such states is being 
considered.

I would like to express my gratitude for the valuable  support in the realization 
of this work of Dr. J.L. Lucio, my colleagues of the Instituto de Fisica of the Guanajuato 
University (Mexico) and Instituto de Cibernetica, Matematica y Fisica (Cuba), 
CONACyT (Mexico) and  the Abdus Salam International Centre for Theoretical 
Physics (Italy).

\newpage

\noi {\large \bf REFERENCES}

\begin{itemize}
\item[1.-] A. Cabo and A. P\'erez Mart\ii nez, Phys. Lett. A 222 (1996) 263.
\item[2.-] A. P\'erez-Mart\ii nez, A. Cabo and V. Guerra. LANL Preprint
        Archives cond-mat/9606222 (1996), appeared in Int. Jour. of Mod. Phys. B.
\item[3.-] A. Cabo and A. P\'erez-Mart\ii nez, LANL Preprint cond-mat/
	   9601019 (1996), to appear in Int. Jour. of Mod. Phys.B
\item[4.-] D. Yoshioka and H. Fukuyama, J. Phys. Soc. Ipn 47 (1979) 394. 
\item[5.-] Y. Kuramoto and R.R. Gerhardts, J. Phys. Soc. Japan 51 (1982) 3810.
\item[6.-] R. Tao and D.J. Thouless, Phys. Rev. B 28 (1983) 1142.
\item[7.-] K. Maki and X. Zotos, Phys. Rev. B 28 (1983) 4349.
\item[8.-] D. Yoshioka and P.D. Lee, Phys. Rev. B 27 (1983) 4986.     
\item[9.-] F. Claro. Sol. State Commun. 53 (1985) 27.
\item[10.-] S.T. Chui, T.M. Hakim and K.B. Ma, Phys. Rev. B 33 (1986) 7110.
\item[11.-] J.S. Kivelson, C. Kallin, D.P. Arovas and J.R. Schrieffer, Phys. 
Rev. Lett. 56 (1986) 873.
\item[12.-] B.I. Halperin, Z. Tesanovic and F. Axel, Phys. Rev. Lett. 57
(1986) 922.
\item[13.-] R. Ferrari, Phys. Rev. B 42 (1990) 4598.
\item[14.-] R. Ferrari, Int. J. Mod. Phys. B 8 (1992)  1992.
\item[15.-] A. Cabo, Phys. Lett. A 171 (1992) 90.
\item[16.-] A. Cabo, Phys. Lett. A 191 (1994) 323.
\item[17.-] A. Cabo, Phys. Lett. A 211 (1996) 297.
\item[18.-] S.M. Girvin, A.H. MacDonald and P.M. Platzman, Phys. Rev. Lett.
54 (1985) 581.
\item[19.-] A.H. MacDonald and S.M. Girvin, Phys. Rev. B 38 (1988) 6295.
\item[20.-] S.M. Girvin, A.H. MacDonald and P.M. Platzman, Phys. Rev. B 33 
(1986) 2481. 
\end{itemize} 
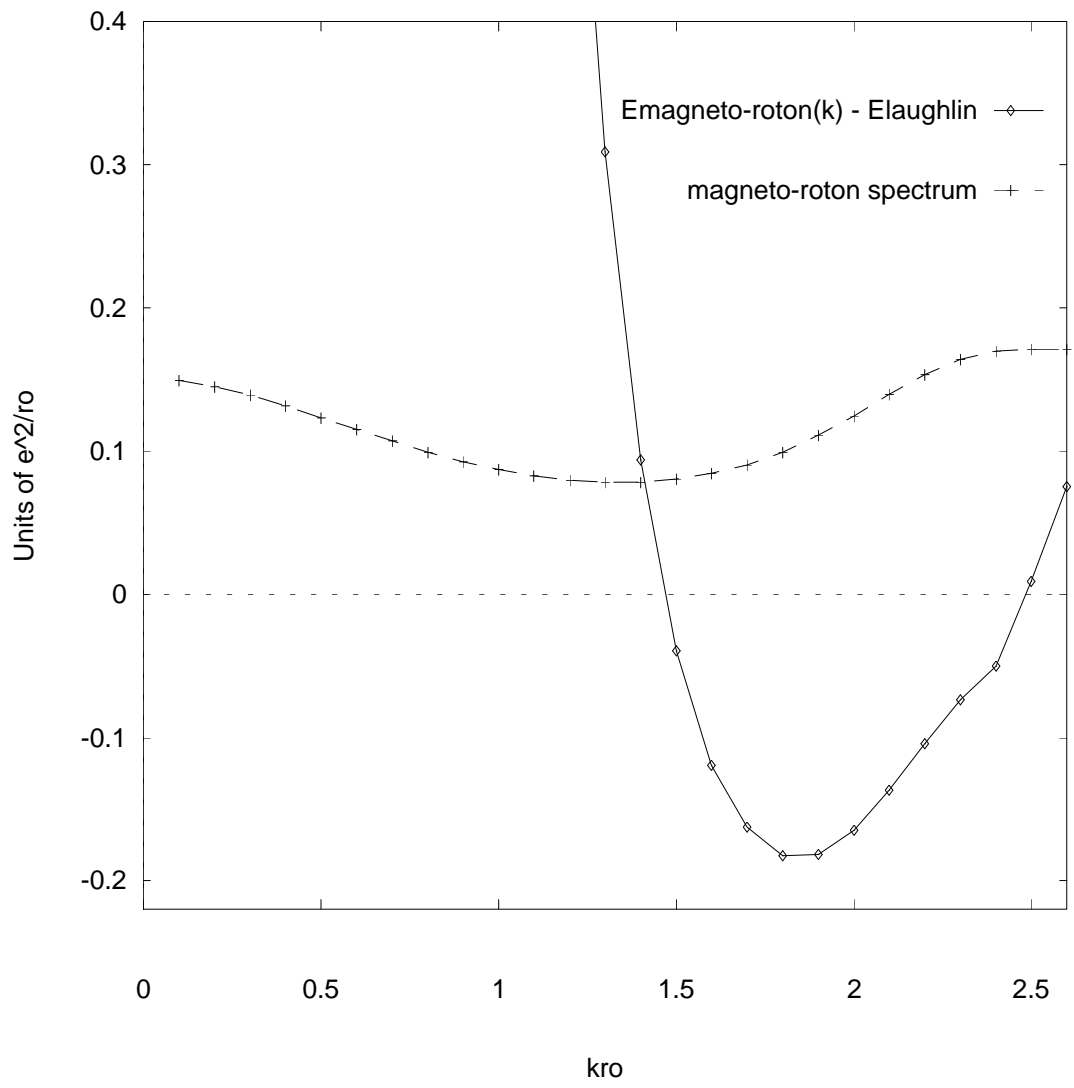
\end{document}